# A non-invasive fault location method for modular multilevel converters under light load conditions

Yaqian Zhang[1], Yi Zhang[2,3], Frede Blaabjerg[2], Jianzhong Zhang[1]
1. School of Electrical Engineering, Southeast University, Nanjing, China
2 AAU Energy, Aalborg University, Aalborg, Denmark
3 Swiss Federal Institute of Technology Lausanne (EPFL), Lausanne, Switzerland
(yaqianzhang83@seu.edu.cn, yiz@energy.aau.dk, fbl@energy.aau.dk, jiz@seu.edu.cn)

*Abstract*—This paper proposes a non-invasive fault location method for modular multilevel converters (MMC) considering light load conditions. The prior-art fault location methods of the MMC are often developed and verified under full load conditions. However, it is revealed that the faulty arm current will be suppressed to be unipolar when the open-circuit fault happens on the submodule switch under light load. This leads to the capacitor voltage of the healthy and faulty submodules rising or falling with the same variations, increasing the difficulty of fault location. The proposed approach of injecting the second-order circulating current will rebuild the bipolar arm current of the MMC and enlarge the capacitor voltage deviations between the healthy and faulty SMs. As a result, the fault location time is significantly shortened. The simulations are carried out to validate the effectiveness of the proposed approach, showing that the fault location time is reduced to 1/6 compared with the condition without second-order circulating current injection.

## I. Introduction

The modular multilevel converter (MMC) is one of the most promising multilevel converters in high-voltage applications thanks to the modularity, superior AC performance, and scalability to meet any voltage level [1]. The half-bridge (HB) submodule (SM) has the simplest structure and lowest power losses, becoming the most popular SM for the MMC. In general, there are even hundreds of identical SMs and massive power switching devices in the MMC, which have a high malfunction rate threatening the reliability of the MMC [2]. Therefore, it is necessary to detect and locate the fault quickly so that further measurements can be taken to protect the MMC from being broken down.

Open-circuit and short-circuit are two typical faults of power semiconductor switches. The short-circuit fault detection is easier and faster to be realized by the switch level circuit, which is usually embedded into the commercial gate driver circuit [3]. However, the open-circuit fault would not cause obvious damage immediately; instead, it remains undetected for a relatively long time and cause a subsequent influence on the overall MMC system [4]. Once an open-circuit fault happens in one or multiple SMs, the output performance of the MMC will deteriorate because the actual output voltage of the faulty SM does not match the required output modes. Moreover, the capacitor in a faulty SM will be overcharged if the faulty SM is not detected and isolated in time. Then the capacitor voltage will be increased to exceed the safe threshold and, at last, break down the whole MMC system [5].

The existing methods of faulty SM location for the MMC are mainly classified into two types, i.e., the hardware-based and software-based methods.

The hardware-based method can realize the fault location within short time. In [6], the voltage sensor is employed to measure the arm voltage and detect the faulty arm. However, it can only be used to locate the defective arm, not the faulty SM. Two voltage sensors are used in each SM in [7] to accurately locate the faulty SM. One voltage sensor is parallel with the capacitor to realize the voltage balancing control, and the other is paralleled to the lower power switch to detect the SM output voltage and compared with the reference siwtching signal. However, the cost is significantly increased since the number of SMs is usually large. In [8], the two voltage sensors are ruduced to one in each SM and the voltage sensor is re-configured to be parallel connected with the upper power switch. The cost is not increased, but the extra voltage observer is required to acquire the capacitor voltage, making the control more complex. Moreover, the above voltage observer will lose the accuracy with the SM capacitance aging, negatively affecting the normal MMC operation. Therefore, with respect to the system cost and operation robustness, the hardware-based approaches of faulty SM location in [7] and [8] are not the best choice.

The software-based methods are proposed for the HBSM MMC without changing the converter structure or sensor configuration. There are three types of software-based methods, namely the signal-based methods, model-based methods, and data-based methods. In the signal-based methods [9], the detected signal is compared with the reference signal, and the fault can be located once the signal deviation exceeds the threshold and last for a certain time. In model-based methods [10], voltage or state observers are demanded to compare the estimated and actual values. Besides, an appropriate threshold is required but difficult to satisfy the requirements under different power levels. In the data-based methods [11], the data feature of the SM capacitor voltage is extracted to detect the error data and derive the error possibility. However, this method depends on the capacitor voltage deviations between the faulty and healthy SMs. Above all, all the aforementioned software-based approaches perform well under the rated power level but are not evaluated under light load conditions.

The arm current will be distorted to be unipolar or even zero by the SM open-circuit faults when the MMC operates under light load [12]. In this case, the capacitor voltage of the healthy and faulty SMs will vary similarly, and the voltage deviations are quite small or even disappear. This will cause the failure of the existing software-based methods since they are all achieved by extracting the capacitor voltage difference between the healthy and faulty SMs.

The fault location of MMC under the light load condition is considered in literature [12], where the hardware-based method is adopted to detect the faulty SM. The bleeding resistor is re-arranged and auxiliary circuits are equipped for each SM to detect the output voltage of the SM in real-time. However, the system costs are increased since the number of SMs is large for the MMC system.

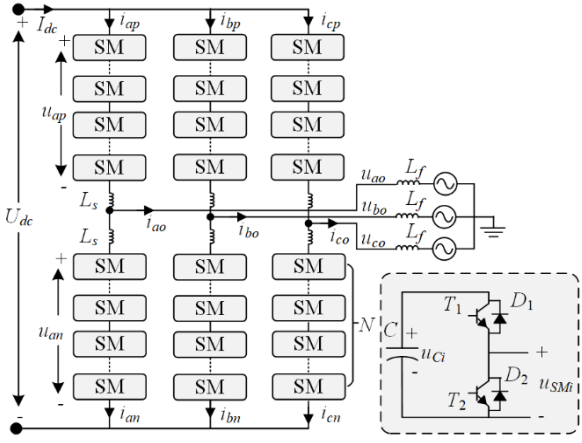

Fig. 1. Configuration of three-phase MMC.

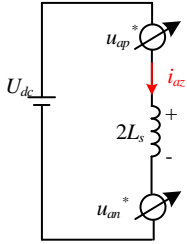

Fig. 2 Equivalent circuit on DC side under normal state.

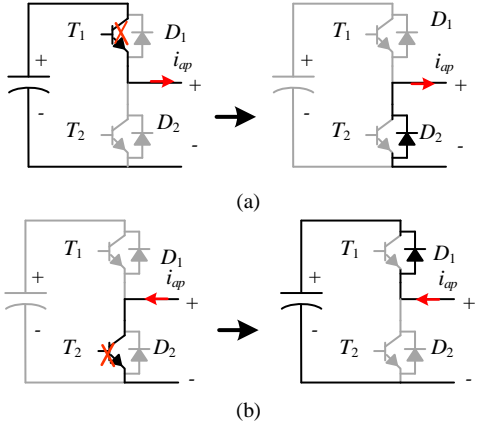

Fig. 3. Open-circuit fault in SM of faulty (a) $T_1$. (b) $T_2$.

In this paper, a software-based fault location method is proposed for the MMC under light load condition to accelerate the fault location speed. First, the mechanism of the unipolar arm current under fault is analyzed and the approximate criterion of high risk of unipolarity is derived. Second, the circulating current injection is enabled after the faulty arm is detected so that the arm current recovers to the bipolar variations. The amplitude and phase of the reference second-order circulating current are selectedfor different types of faults. Then the capacitor voltage deviations between the faulty and healthy SMs are appropriately enlarged to shorten the fault location time.

## II. OPEN-CIRCUIT FAULT OF MODULAR MULTILEVEL CONVERTERS

### A. Configuration of MMC

Table I. States of Healthy SM

| Mode | $i_{ap}$ | $S_{api}$ | State | Path | $u_{SMi}$ | $u_{Ci}$ |
|------|----------|-----------|--------|-------|-----------|----------|
| 1 | ≥0 | 1 | Insert | $D_1$ | $+u_{Ci}$ | Increase |
| 2 | ≥0 | 0 | Bypass | $T_2$ | 0 | Unchanged |
| 3 | <0 | 1 | Insert | $T_1$ | $+u_{Ci}$ | Decrease |
| 4 | <0 | 0 | Bypass | $D_2$ | 0 | Unchanged |

Table II. Two types of Open-circuit Fault

| Fault type | Mode | $i_{ap}$ | $S_{api}$ | State | $u_{SMi}$ | $u_{Ci}$ |
|------------|------|----------|-----------|--------|-----------|----------|
| $T_1$ fault | 1 | ≥0 | 1 | Insert | $+u_{Ci}$ | Increase |
| | 2 | ≥0 | 0 | Bypass | 0 | Unchanged |
| | 3 | <0 | 1 | Bypass | 0 | Unchanged |
| | 4 | <0 | 0 | Bypass | 0 | Unchanged |
| $T_2$ fault | 1 | ≥0 | 1 | Insert | $+u_{Ci}$ | Increase |
| | 2 | ≥0 | 0 | Insert | $+u_{Ci}$ | Increase |
| | 3 | <0 | 1 | Insert | $+u_{Ci}$ | Decrease |
| | 4 | <0 | 0 | Bypass | 0 | Unchanged |

Fig. 1 shows the configuration of the three-phase MMC, where each phase consists of upper and lower arms. Each arm contains one arm inductor and $N$ identical HBSMs. In each HBSM, $u_{Ci}$ represents the capacitor voltage. $u_{SMi}$ is the output voltage of the SM, having $+u_{Ci}$ and 0, namely the insert and bypass states, respectively. $i$ is the numbering of each SM in each arm, having $i=1,…,N$. $u_{xy}$ denotes the summation of HBSM output voltage in the $y$ arm of phase $x$, where $x=a$, $b$, $c$, $y=p, n$.

Since the HBSM can only output nonnegative voltage, the charging or discharging states of the SM capacitor are determined by the arm current, which is shown in Table I. When the arm current is positive, the SM is inserted via $D_1$ and bypassed via $T_2$. When the arm current is negative, the SM is inserted by $T_1$ and bypassed by $D_2$.

The arm voltage and current of each arm are expressed as

$$\begin{cases} u_{xy} = \dfrac{U_{dc}}{2} \mp U_m \sin(\omega t + \varphi_x) \\ i_{xy} = \dfrac{I_{dc}}{3} \pm I_m \sin(\omega t + \varphi_x + \varphi) \end{cases} \quad (1)$$

where only DC and fundamental components are included in the arm currents with the higher-order currents suppressed. According to the power balancing between the DC and AC sides, the DC and AC current have the relations of

$$I_{dc} = \dfrac{3m}{4} I_m \cos\varphi \quad (2)$$

where $\varphi$ is the power factor angle of the MMC. $m$ is the modulation index of the MMC, having $U_m/(U_{dc}/2)$.

Taking phase $a$ as an example and seeing from the DC side, the equivalent circuit of the MMC can be illustrated in Fig. 2. We have

$$U_{dc} - (u_{ap} + u_{an}) = 2L_s \dfrac{di_{az}}{dt} \quad (3)$$

where $i_{az}$ is the circulating current between the DC side and phase $a$. Under the normal operations in Fig. 3, the actual $u_{ap}$ and $u_{an}$ follow the required reference value $u_{ap}^*$ and $u_{an}^*$, respectively. $u_{ap}^*$ and $u_{an}^*$ have

$$U_{dc} - \left(u_{ap}^* + u_{an}^*\right) = 0 \tag{4}$$

Then the right term of (1) is zero without fault occurrence, which is

$$2L_s \frac{di_{az}}{dt} = 0 \tag{5}$$

which is consistent with the fact that the circulating current is controlled as the DC component under normal operation.

*B. Open-circuit Fault*

Fig. 3 shows two types of open-circuit faults on $T_1$ and $T_2$. In Fig. 3(a), the SM cannot be inserted under $i_{ap}<0$ when the upper switch $T_1$ is faulty, and the arm current can only flow through the freewheeling diode $D_2$. Therefore, as shown in Table II, the faulty SM cannot be discharged in Mode 3, leading to over-voltage of the SM capacitor. In Fig. 3(b), the SM cannot be bypassed by $T_2$ under $i_{ap}>0$ with open-circuit occurrence on $T_2$. In this case, the arm current flows via the freewheeling diode $D_1$ and charges the SM capacitor. As a result, the capacitor voltage of the faulty SM keeps increasing as shown in mode 2 of $T_2$ fault in Table II. This will also cause over-voltage of the faulty SM.

Based on the above analysis, the open-circuit fault of $T_1$ or $T_2$ will both cause an abnormal increase and divergence of the corresponding SM capacitor voltage. Also, as the faulty SM cannot output the voltage required by the switching signal, the arm voltage is distorted, causing circulating and arm current distortion. This fault of arm current or circulating current is used to detect the faulty arm or phase. But the specific faulty SM must be located by distinguishing its own capacitor voltage or other information inside each SM.

## III. IMPACT OF LIGHT LOAD ON IGBT OPEN-CIRCUIT FAULT LOCATION

This section clarifies the impact of the system parameters and operation conditions on the faulty arm current polarities. It emphasizes the negative impact of neglecting the light load condition on the fault location of the MMC.

*A. Unipolar arm current under $T_1$ fault*

Assume that the faulty SM exist in the upper arm. Under the open-circuit fault of $T_1$, SM1 cannot be inserted but only be bypassed by the diode $D_2$ with $i_{ap}<0$. Then the actual arm voltage is smaller than the reference arm voltage, which is

$$u_{ap\_f1} = \begin{cases} u_{ap}^* - S \cdot U_C & i_{ap} < 0 \\ u_{ap}^* & i_{ap} > 0 \end{cases} \tag{6}$$

where $S$ is the reference switching function of the faulty SM, having $S=0,1$.

Fig. 4 illustrates the equivalent circuit under $T_1$ fault. The voltage deviation under $T_1$ fault is given by

$$\Delta u_{ap\_f1} = u_{ap}^* - u_{ap\_f1} = \begin{cases} S \cdot U_C & i_{ap} < 0 \\ 0 & i_{ap} > 0 \end{cases} \tag{7}$$

Here we assume that the faulty SM capacitor voltage is not changed too much and is still limited to the safety zone without over-voltage before it sis located. In Fig. 4(c), $\Delta u_{ap\_f1}$ is evenly distributed on the two arm inductors. Then the faulty current component under $T_1$ is

$$\Delta i_{ap\_f1} = \int \frac{S \cdot U_c}{2L_s} dt \quad i_{ap} < 0 \tag{8}$$

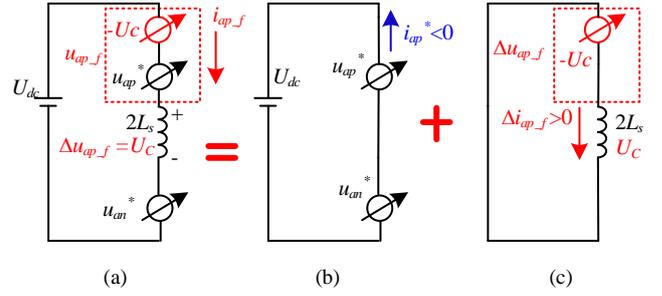

Fig. 4. Equivalent circuit of $T_1$ fault. (a) Overall circuit. (b) Normal part. (c) Faulty part.

In Fig. 4(a), the faulty arm current $i_{ap\_f1}$ is composed of the normal component $i_{ap}^*$ and the faulty component $\Delta i_{ap\_f}$, which is given as

$$i_{ap\_f1} = \Delta i_{ap\_f1} + i_{ap}^* \quad i_{ap\_f1} < 0 \tag{9}$$

Since the faulty component in (8) is positive, the negative part of the arm current becomes smaller under the open-circuit fault. In particular, the negative part of $i_{ap\_f1}$ might disappear under light load.

If the following prerequisite is satisfied, the negative part of the arm current will be canceled out, leading to the unipolar faulty arm current with only the nonnegative part.

$$\Delta i_{ap\_f} \geq \max\left(\left|i_{ap}^*\right|\right) \quad i_{ap} < 0 \tag{10}$$

Combining (10) with (1), we have

$$\max\left(\left|i_{ap}^*\right|\right) = \frac{I_m}{2} - \frac{I_{dc}}{3} \quad i_{ap} < 0 \tag{11}$$

Combining (2), (8) and (11), the criterion in (10) is given as

$$\int (S) dt \geq \left(\frac{1}{2} - \frac{m\cos\varphi}{4}\right) \frac{2I_m L_s}{U_c} \tag{12}$$

To make sure that (12) is satisfied, the left term is suppressed as the integration of one switching period, which is approximated as

$$\int (S) dt = \left(\frac{1}{2} - \frac{m}{2}\sin(\omega t)\right) \cdot \Delta T_c \approx \frac{1}{2f_c} \tag{13}$$

where $f_c$ is the carrier and switching frequency.

Combining (12) and (13), the critical condition to generate the unipolar arm current is expressed as

$$I_m = \frac{U_{dc}}{(2 - m\cos\varphi)} \cdot \frac{1}{Nf_c L_s} \tag{14}$$

In case of the multiple faulty SMs, $N_{f1}$ rfers to the number of SMs with $T_1$ fault. Then (14) is written as

$$I_m = \frac{U_{dc} \cdot N_{f1}}{(2 - m\cos\varphi)} \cdot \frac{1}{Nf_c L_s} \tag{15}$$

which means that, if the AC current amplitude is lower than the right side, the faulty arm current will become nonnegative. As a result, all the SMs cannot be discharged and have the same voltage variations as the faulty SM1. This will cause the failure to locate the faulty SM quickly.

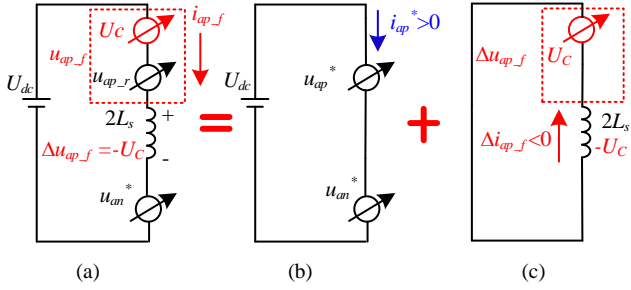

Fig. 5 Equivalent circuit of $T_2$ fault. (a) Overall circuit. (b) Normal part. (c) Faulty part.

Table III Cases of different impact factors

| Impact factors | Case | Parameters |
|---|---|---|
| Arm inductance $L_s$ | 1 | $L_s$=8mL, $\cos\varphi$=1, $f_c$=1.2 kHz, $m$=0.9 |
| | 2 | $L_s$=5mL, $\cos\varphi$=1, $f_c$=1.2 kHz, $m$=0.9 |
| Modulation index $m$ | 3 | $L_s$=8mL, $\cos\varphi$=1, $f_c$=2 kHz, $m$=0.9 |
| | 4 | $L_s$=8mL, $\cos\varphi$=1, $f_c$=2 kHz, $m$=0.8 |
| Power factor $\cos\varphi$ | 5 | $L_s$=8mL, $\cos\varphi$= 1, $f_c$=1.2 kHz, $T_1$ |
| | 6 | $L_s$=8mL, $\cos\varphi$=0.86, $f_c$=1.2 kHz, $T_2$ |
| Number of faulty SMs $N_{f1}$, $N_{f2}$ | 7 | $L_s$=8mL, $\cos\varphi$=0, $f_c$=1.2 kHz, $N_{f1}$=1 |
| | 8 | $L_s$=8mL, $\cos\varphi$=0, $f_c$=1.2 kHz, $N_{f1}$=2 |
| | 9 | $L_s$=8mL, $\cos\varphi$=0, $f_c$=1.2 kHz, $N_{f2}$=3 |

### B. Unipolar arm current under $T_2$ fault

If $T_2$ suffers from the open circuit fault, SM1 cannot be bypassed under $i_{ap}>0$ and then the actual arm voltage $u_{ap}$ becomes

$$u_{ap\_f2} = \begin{cases} u_{ap}^* + S \cdot U_C & i_{ap} > 0 \\ u_{ap}^* & i_{ap} < 0 \end{cases} \quad (16)$$

As seen from the equivalent circuit in Fig. 5, the faulty arm voltage deviation under $T_2$ fault is

$$\Delta u_{ap\_f2} = u_{ap}^* - u_{ap\_f2} = \begin{cases} -(1-S) \cdot U_C & i_{ap} > 0 \\ 0 & i_{ap} < 0 \end{cases} \quad (17)$$

which is imposed on the arm inductors and generates the faulty component of the arm current:

$$\Delta i_{ap\_f2} = \int \frac{(S-1) \cdot U_c}{2L_s} dt \quad i_{ap} > 0 \quad (18)$$

which is negative under the positive arm current. It means that the faulty component is always opposite to the normal component, then the positive interval of the arm current will be cut off, even causing unipolarity.

Similar to the analysis of the $T_1$ fault, the unipolar arm current under $T_2$ fault occurs under the following conditions:

$$I_m = \frac{U_{dc} \cdot N_{f2}}{(2+m\cos\varphi)} \cdot \frac{1}{Nf_c L_s} \quad (19)$$

where $N_{f2}$ is the number of SMs with $T_2$ fault. It means that when the AC current amplitude is lower than the right term, then there will be high possibility of nonpositive arm current.

In (15) and (19), it shows that whether the faulty arm current will become unipolar is affected by various factors, i.e., the number of SMs when the DC bus voltage is fixed, the power/ current level, the arm inductance, the carrier frequency, and the power factor. To emphasize, the cretirion in (15) and (19) are derived with certain simplification, and only the dominant impacts are included. There might be some errors compared to the actual operations since the control strategies, such as the circulating current control and the voltage balancing control, will also affect the faulty phenomenon. It provides a clue that when the unipolar arm current occurs and more measures will be taken to accelerate the fault location or avoid possible mislocation.

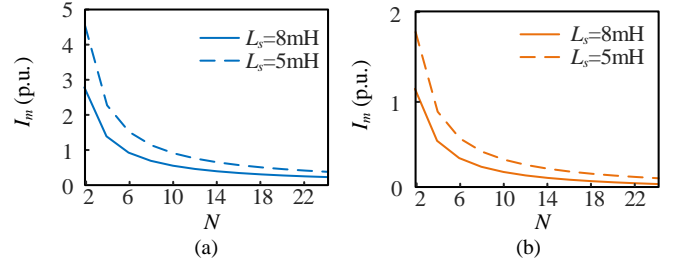

Fig. 6 Impact of arm inducatance on faulty arm current. (a) $T_1$ fault. (b) $T_2$ fault. (Case 1 and case 2).

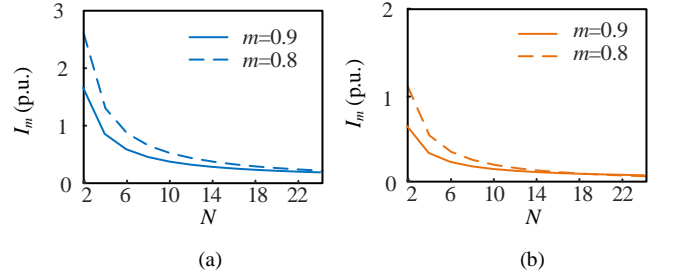

Fig. 7 Impact of modulatioin index on faulty arm current. (a) $T_1$ fault. (b) $T_2$ fault. (Case 3 and case 4).

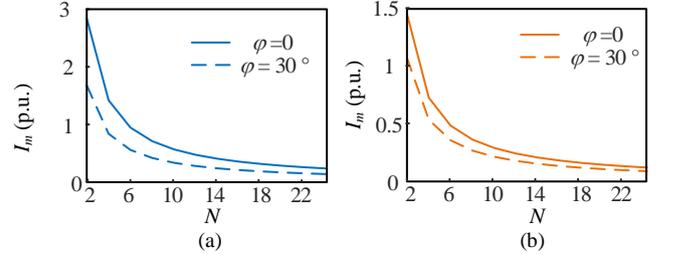

Fig. 8 Impact of power factor on faulty arm current. (a) $T_1$ fault. (b) $T_2$ fault. (Case 5 and case 6).

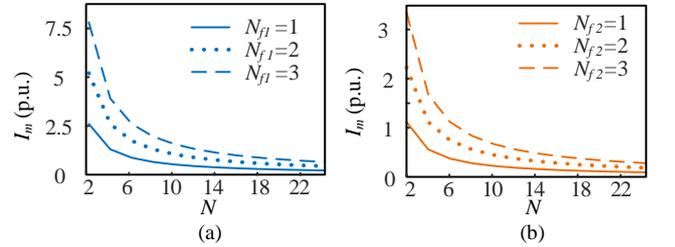

Fig. 9 Impact of number of faulty SMs on faulty arm current. (a) $T_1$ fault. (b) $T_2$ fault. (Cases 7-9).

### C. Impact factor of unipolar faulty arm current

According to the criterion in (15) and (19), the impact factors are analyzed in detail by drawing the curves to illustrate the tendency, as shown in Figs. 6-9, and the corresponding parameters of each case are shown in Table III.

Firstly, the curves of the $I_m$-$N$ are drawn for different arm inductance in Fig. 6. All of the points on the curves indicate the critical operation condition of the unipolar faulty arm current. Under the given number of SMs, the corresponding value of $I_m$ indicates that the unipolar arm current could occur under smaller current or lower power. The higher $I_m$ threshold indicates a higher risk of unipolar faulty arm current. With the increase of the SM quantity under the fixed DC bus voltage, the $I_m$ threshold decreases, meaning that higher SM voltage is easier to trigger the unipolar arm current. The $I_m$ threshold of $T_1$ and $T_2$ fault are both increased when the arm inductance is decreased from 8 mH to 5 mH. Comparing Fig. 6(a) and Fig. 6(b), it shows that the $I_m$ threshold of $T_1$ fault is higher than $T_2$. In other words, the faulty arm current of $T_1$ fault is easier to be unipolar under the same AC current level.

Secondly, the curves of the $I_m$-$N$ are compared under $m=0.9$ and $m=0.8$ in Fig. 7. Even though the decrease of modulation index will increase the $I_m$ threshold, the impact is slight, especially in the case of the high number of SMs. Note that the carrier frequency in case 3 and case 4 is increased to 2 kHz compared to case 1 and case 2. In Table III, case 1 and case 3 can be used to compare the influence of carrier frequency, which are curved as the solid lines in Fig. 6 and Fig. 7. Obviously, the increase of carrier frequency reduces the $I_m$ threshold decreases the $I_m$ threshold.

Thirdly, the influence of the power factor is displayed in Fig. 8. It shows that the reactive power can help to reduce $I_m$ threshold and reduce the unipolar risk under faulty $T_1$ or $T_2$.

Finally, the multiple SM faults with the same type of fault are shown to have high risk of causing unipolar arm current, as seen in Fig. 9 where higher number of faulty SMs produces higher $I_m$ threshold.

Above all, the increase of faulty SM quantity and power factor will increase the risk of unipolar arm current; however, the increase of modulation index, carrier frequency, and arm inductance will decrease the risk of unipolar arm current. Under the fixed DC bus voltage, the decrease of SM quantity will increase the SM voltage, finally causing higher risk of unipolar arm current.

## IV. PROPOSED FAULT LOCATION METHOD

### A. Mechanism of faulty location

The existing software-based fault location methods are classified as three types, including the signal-, model-, and data-based methods. Basically, all of the methods are realized by distinguishing the faulty SM capcitor voltage with the healthy SM capacitor voltage. However, under the unipolar faulty arm current, the faulty and healthy SM capacitor voltage will rise or fall similarly. For example, the faulty arm current becomes nonnegative under $T_1$ fault, then all of the SM capacitors cannot be discharged and the faulty SM is hiden. Similarly, $T_2$ fault cannot be located under non-positive faulty arm current.

Therefore, to ensure that all of the existing approach can be well used without being degraded by the faulty unipolar arm current, the point is to produce bipolar arm current under $T_1$ or $T_2$ fault.

### B. Proposed solution to unipolar arm current

This paper proposes the second-order circulating current injection approach as a solution to unipolar faulty arm current. The injected circulating current can rebuild the arm current to be bipolar so that the capacitor voltage deviations between the healthy and false SMs will be generated or enlarged during $i_{ap\_f2}>0$ or $i_{ap\_f1}<0$.

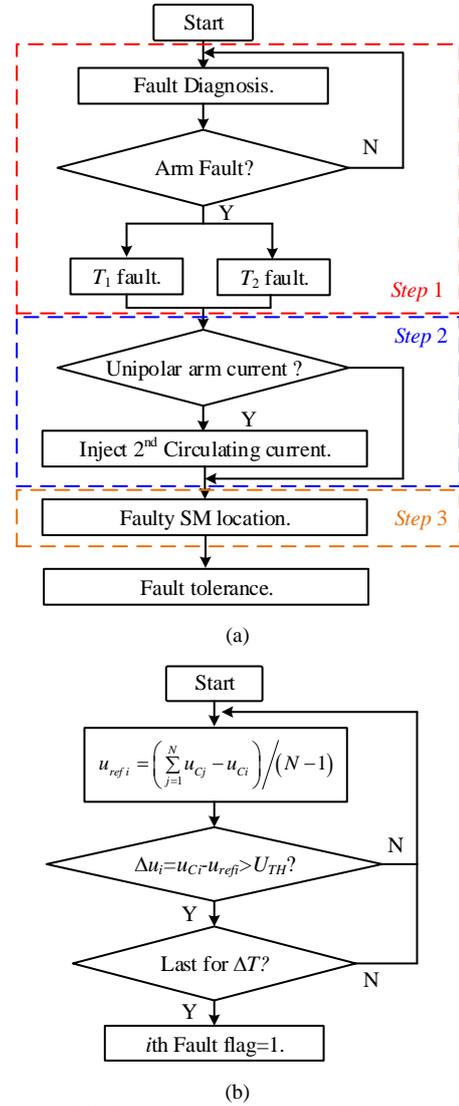

Fig. 10 Proposed fault location method under light load. (a) Enable of circulating current injection. (b) Fault location method.

Fig. 10 shows the implementation of the proposed fault location methods based on the circulating current injection. In Fig. 10(a), the fault location approach is performed in three steps.

***Step 1:*** The fault diagnosis is performed to detect the faulty arm and type of faults. The approach in [13] can be directly used.

***Step 2:*** The circulating current injection is only enabled after the arm fault is detected and the criterion in (15) or (19) is satisfied. The injected circulating current is given as

$$i_{2nd\_x} = I_{2nd} \sin\left(\omega t + \varphi_x + \varphi_{2nd}\right) \quad (20)$$

where $I_{2nd}$ and $\varphi_{2nd}$ are the amplitude and phase of the second order circulating current, respectively. Specifically, the phase is determined by the fault type, i.e., $T_1$ or $T_2$ type, which is

$$\varphi_{2nd} = \begin{cases} -\dfrac{\pi}{2} & T_1 \ fault \\ \dfrac{\pi}{2} & T_2 \ fault \end{cases} \quad (21)$$

In this way, the missing negative/ positive interval of the faulty arm current could be compensated to the maximum extent. The amplitude of the injected second circulating current is decided by the current rating of the power device.

Table IV Simulation Parameters of MMC

| Parameters | Value |
|---|---|
| DC voltage ($U_{dc}$) | 24 kV |
| AC Voltage Amplitude ($U_m$) | 11 kV |
| Number of SMs ($N$) | 12 |
| Nominal Power ($S$) | 7.5 MVA |
| SM Capacitance ($C$) | 3 mF |
| Nominal SM Voltage ($U_C$) | 2.0 kV |

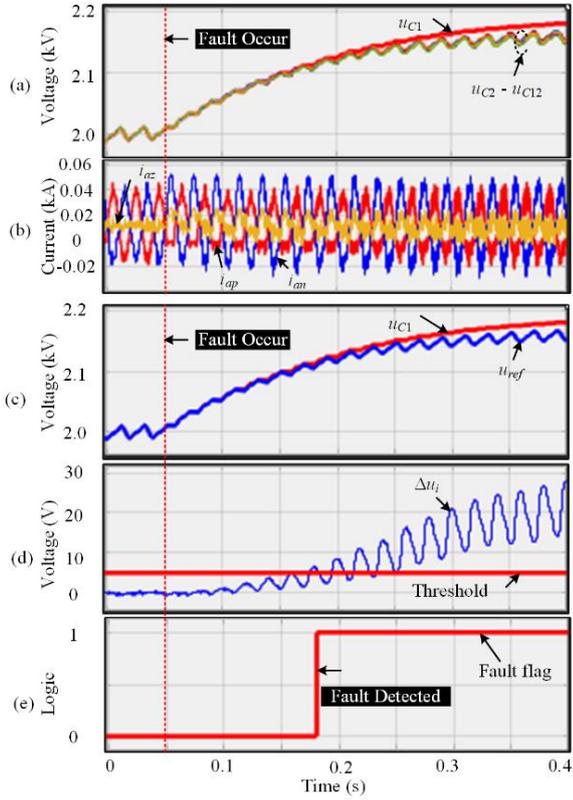

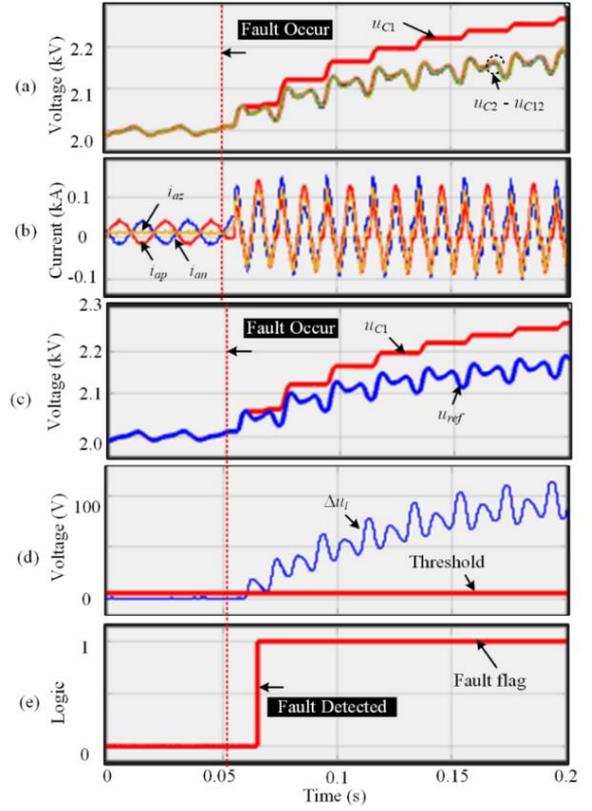

Fig. 11. Faulty SM location under $T_1$ fault and light load, without second order circulating current injection. (a) SM capacitor voltage. (b) Arm currents and circulating current. (c) SM1 capacitor voltage and reference

Then the circulating current control in [14] is adopted to inject the second order circulating current; however, the injected circulating current cannot follow the reference value due to the false switching of the faulty SM. Nevertheless, the point is producing bipolar arm current rather than accurate circulating current track. Note that the healthy phase should also be compensated by the circulating current so that the DC link current will not be too severely distorted.

***Step 3:*** After the circulating current is injected, the fault location is conducted, as shown in Fig. 10(b), where the classic method in [4] is adopted. The capacitor voltage of the detected SM is compared with the average voltage of the rest SMs $u_{refi}$, and the calculated voltage deviation $\Delta u_i$ is compared with the given threshold $U_{TH}$. If $\Delta u_i > U_{TH}$ lasts for $\Delta T$, then SM$i$ can be detected as faulty SM and bypassed by the outer bypass breaker. $\Delta T$ is set as 5 ms in this paper.

Fig. 12. Faulty SM location under $T_1$ fault and light load, with second-order circulating current injection. (a) SM capacitor voltage. (b) Arm currents and circulating current. (c) SM1 capacitor voltage and reference value. (d) Comparison with threshold. (e) Fault flag of SM1.

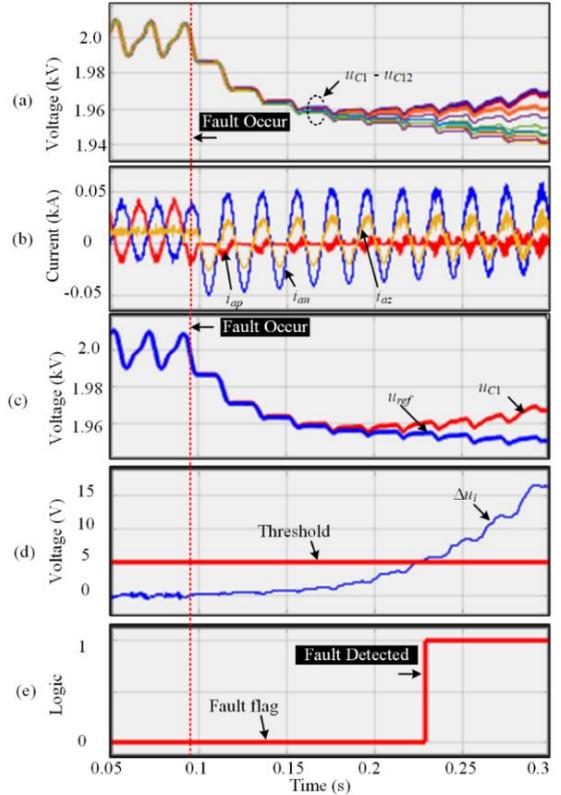

Fig. 13. Faulty SM location under $T_2$ fault and light load, without second order circulating current injection. (a) SM capacitor voltage. (b) Arm currents and circulating current. (c) SM1 capacitor voltage and reference value.(d) Comparison with threshold. (e) Fault flag of SM1.

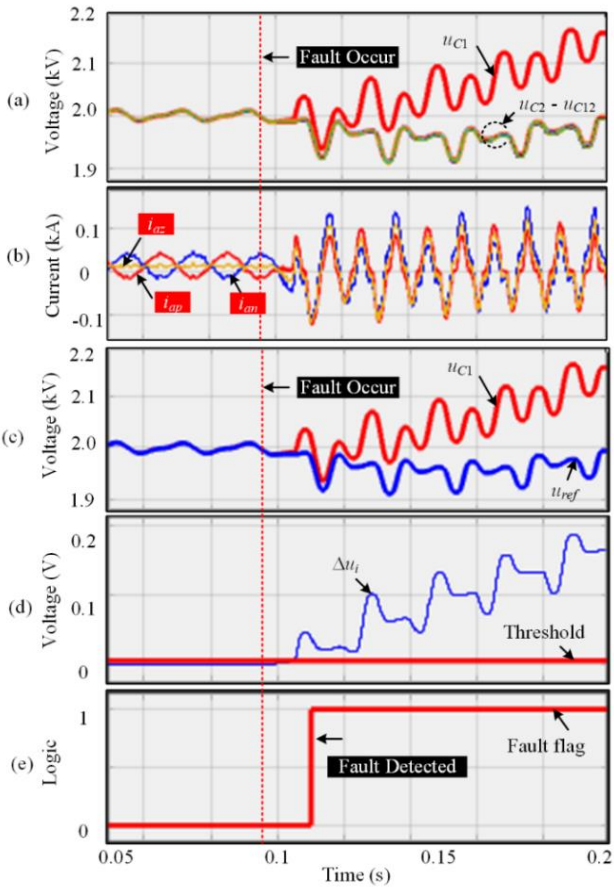

Fig. 14. Faulty SM location under $T_2$ fault and light load, with second order circulating current injection. (a) SM capacitor voltage. (b) Arm currents and circulating current. (c) SM1 capacitor voltage and reference value. (d) Comparison with threshold. (e) Fault flag of SM1.

Table V Fault location time under light load condition

| Fault case | Circulating current injection | |
|---|---|---|
| | Without | With |
| $T_1$ fault | 130 ms | 17 ms |
| $T_2$ fault | 135 ms | 20 ms |

## V. VERIFICATIONS

The simulation verifications are carried out in Simulink/ Matlab on a three-phase MMC with the parameters shown in Table IV. The carrier phase shift-based pulse width modulation approach is adopted to realize the voltage balancing for individual SM. The SM1 in the upper arm of phase *a* is assumed to be faulty. The simulation results are presented for the $T_1$ and $T_2$ fault under the power level $P$= 750 kW (0.1 p.u.) with $I_m$= 50 A. According to the given criterion in (15) and (19), the arm current will suffer from unipolarity under both $T_1$ and $T_2$ faults.

### A. $T_1$ fault

Fig. 11 shows the SM fault location process under $T_1$ fault, where no extra measurement is taken. After the fault occurrence at $t$= 0.04s, the SM capacitor voltage in Fig. 11(a) is almost the same before $t$=0.16 s. The arm current of the fault upper arm in Fig. 11(b) is nearly unipolar, having only positive intervals. As a result, the voltage deviation between $u_{C1}$ and $u_{ref1}$ increases slowly and finally exceeds the threshold at $t$= 0.17 s, where the threshold value is 5 V. Lasting for 10 ms, the fault flag of SM1 is turned to one, and the faulty SM is detected. The fault location time is 130 ms.

In Fig. 12, the proposed approach of second-order circulating current is employed for the fault location of $T_1$. The fault occurs at $t$=0.05 s, and the detection time of the faulty arm is assumed to be 10 ms. Therefore, the circulating current injection is enabled at 0.06 s. It is shown in Fig. 12(a) that the voltage deviations between the healthy and faulty SM capacitors are enlarged by the circulating current injection. The voltage deviation $\Delta u_i$ in Fig. 12(d) rises quickly, and the fault is located at $t$=0.67 s. Compared with Fig. 11, the fault location process is accelerated, and the location time is significantly reduced from 130 ms to 17 ms.

### B. $T_2$ fault

Fig. 13 and Fig. 14 show the fault location under $T_2$ fault and light load when the circulating current injection is disabled and enabled, respectively.

It is shown in Fig. 13 that, after the fault happens at $t$=0.95 s, the positive interval of the faulty arm current is suppressed to almost zero, and the SM capacitor voltage is all decreased without apparent deviations. Then $\Delta u_1$ rises quite slowly and cannot reach the threshold until $t$=0.22 s. As a result, the faulty SM is located at $t$=0.23 s. The fault location time is 135 ms.

In Fig. 14, the second-order circulating current is injected at $t$=0.1 s, and the arm current becomes bipolar. Then the faulty SM capacitor voltage rises fast because it cannot be bypassed under positive arm current. As a result, $\Delta u_1$ rises quickly to surpass the threshold, and the fault is located at $t$=0.11 s. The fault location time is shortened to 20 ms.

Table V compares the fault location time with and without circualting current injection for the two types of faults. It justifies the effectiveness of the proposed approach of second-order circulating current injection.

## VI. CONCLUSION

This paper proposes a non-invasive fault location method for MMC to accelerate the fault location process under the light load operation. The arm current is suppressed to be unipolar when the open-circuit fault happens on the SM switch under light load. The approach of injecting the second-order circulating current will rebuild the bipolar arm current and enlarge the capacitor voltage deviations between the healthy and faulty SMs. As a result, the fault location time is significantly shortened. The simulations are carried out to validate the effectiveness of the proposed approach, presenting that the fault location time is reduced to 1/6 compared with the condition without using the second-order circulating current injection.

Also, considering possible negative impacts, the second order circulating current should be injected not surpassing five fundamental periods, which are sufficient to locate the faulty SMs by the proposed approach. Note that the proposed approach is only an improvement before the implementation of the fault location, so it is suitable for all of the existing fault location principles.